\documentclass[%
reprint,
amsmath,amssymb,
prl,
]{revtex4-1}

\usepackage{graphicx}
\usepackage{dcolumn}
\usepackage{bm}

\usepackage[margin=0.73in]{geometry}

\begin{document}


\title{Dynamic induced softening in frictional granular material\\
	investigated by DEM simulation}

\author{Laure Lemrich}
  \email{laure.lemrich@alumni.epfl.ch}
  \affiliation{Chair of Building Physics, ETHZ, Wolfgang-Paulistrasse 15, 
  CH-8093 Zurich, Switzerland\\
  	Laboratory of Multiscale Studies in Building Physics, Empa, 
  	\"Uberlandstrasse 129, CH-8600 D\"ubendorf, Switzerland}

\author{Paul A. Johnson}
  \affiliation{Solid Earth Geophysics Group, Los Alamos National Laboratory, MS 
  D443, Los Alamos, 87545 New Mexico, USA}

\author{Robert Guyer}
  \affiliation{Solid Earth Geophysics Group, Los Alamos National Laboratory, MS 
  D443, Los Alamos, 87545 New Mexico, USA\\
  Department of Physics, University of Nevada, Reno (NV), USA	}

\author{Xiaoping Jia}
  \affiliation{Institut Langevin, ESPCI Paris, CNRS UMR 7587 - 1 rue Jussieu, 
  75005 Paris, France}

\author{Jan Carmeliet}
  \affiliation{Chair of Building Physics, ETHZ, Wolfgang-Paulistrasse 15, 
  CH-8093 Zurich, Switzerland\\
  	Laboratory of Multiscale Studies in Building Physics, Empa, 
  	\"Uberlandstrasse 129, CH-8600 D\"ubendorf, Switzerland}

\date{\today}

\begin{abstract}
A granular system composed of frictional glass beads is simulated using the 
Discrete Element Method. The inter-grain forces are based on the Hertz contact 
law in the normal direction with frictional tangential force. The damping due 
to collision is also accounted for. Systems are loaded at various stresses and 
their quasi-static elastic moduli are characterized. Each system is subjected 
to an extensive dynamic testing protocol by measuring the resonant response to 
a broad range of AC drive amplitudes and frequencies via a set of diagnostic 
strains. The system, linear at small AC drive amplitudes has resonance 
frequencies that shift downward (i.e., modulus softening) with increased AC 
drive amplitude. Detailed testing shows that the slipping contact ratio does 
not contribute significantly to this dynamic modulus softening, but the 
coordination number is strongly correlated to this reduction. This suggests 
that the softening arises from the extended structural change via break and 
remake of contacts during the rearrangement of bead positions driven by the AC 
amplitude.
\end{abstract}


\maketitle


\section{Introduction}

Granular materials are comprised of an ensemble of randomly packed solid 
particles and the mechanical behaviour of the systems is basically determined 
by the interactions at contacts. These materials are ubiquitous in industry and 
in geosciences, and are also of fundamental interest to ground motion and 
earthquake dynamics. Unlike ordinary materials, granular media can exhibit 
solid-like and fluid-like behaviour and there exists transition between the two 
states \cite{liu1998nonlinear}. A granular solid shows strong nonlinear 
elasticity and sound propagation provides a footprint of this feature 
\cite{goddard1990nonlinear, liu1993sound,jia1999ultrasound,makse2004granular}.
The nonlinear dynamic response found in granular media such as resonance 
frequency softening, slow dynamics and harmonic generation 
\cite{norris1997nonlinear,johnson2005nonlinear,brunet2008,jia2011elastic} is 
very similar to those discovered in rocks \cite{johnson1996resonance, 
guyer1999nonlinear, smith2000sensitive, 
ostrovsky2001dynamic,tencate2004nonlinear}.

Other nonlinear behavior observed in granular solids include stress-strain 
hysteresis \cite{ ostrovsky2001dynamic}, fabric anisotropy 
\cite{johnson1998linear, khidas2010anisotropic} and loading-history-dependent 
sound velocity 
\cite{norris1997nonlinear,johnson1998linear,khidas2010anisotropic}. Such 
behaviour is closely related to the fragility of the granular solid determined 
by the very inhomogeneous and anisotropic contact network ; it may react 
elastically to load changes in one specific direction but infinitesimal loads 
in another direction will drive rearrangements in the sample  
\cite{cates1998jamming}. 

If the jammed granular solids are treated as homogenous, for example by coarse 
graining \cite{goldenberg2005friction}, the effective medium theory (EMT)  
\cite{digby1981effective,walton1987effective,makse1999effective}  may be 
applied using an affine approximation to qualitatively connect the global 
response, like the bulk and shear elastic moduli $K$ and $G$, to the local 
geometry. In the case of isotropic compression, the EMT based on the Hertz 
contact law predicts the scaling for  $K \propto (\phi Z_c)^{2/3}\sigma^{1/3}$ 
and $G\propto(\phi Z_c)^{2/3}\sigma^{1/3}$ where $\sigma$ is the confining 
pressure, $\phi$ the packing fraction and $Z_c$ the coordination number 
(assumed to be 
constant)\cite{digby1981effective,walton1987effective,makse2004granular}. The 
scaling of $K$ and $G$ can be determined by the velocities of compression and 
shear waves, via $v_{P}=[(K+2G)/\rho]^{1/2}$ and $v_{S}=(G/\rho)^{1/2}$ where 
$\rho=\rho_{0}\phi$ and $\rho_{0}$ are respectively the packing and the 
particle densities \cite{digby1981effective,makse2004granular}. It has been 
revealed by numerical simulations that at low $\sigma$ the bulk modulus still 
scales as $K \propto \sigma^{1/3}$ while the shear modulus scales as $G \propto 
\sigma^{2/3}$ near unjamming \cite{wyart2005effects, makse2004granular}. This 
observation can be explained by the breakdown of the effective medium theory 
due to the nonaffine motion caused by the rearrangement of the contact network 
or particle positions.	
\newline%

Current numerical works on vibrational properties focus on the density of modes 
as the packing fraction decreases to the jamming point \cite{OHern2003jamming,Mouraille2005sound,wyart2005effects, 
somfai2005elastic, vitelli2010heat, xu2010anharmonic, reichhardt2015softening}. 
An issue of broad interest is the evolution of material characteristics as a 
granular packing approaches the unjamming transition by vibration-induced 
fluidization \cite{dAnna2001jamming}. The combination of acoustic probing and 
pumping in the nonlinear regime allows us to highlight the material softening 
near this transition, using sound velocity \cite{jia2011elastic,wildenberg2013} 
or resonance \cite{johnson2005nonlinear} measurements. Simulations with 2D disk 
packings using a normal Hertzian contact force and a tangential viscous force 
have qualitatively reproduced the experimental observation of resonance 
frequency softening with increasing dynamic amplitude. This behaviour arises 
from rearrangements of the contact network, resulting in a reduction in the 
average contact number but without significant rearrangement of particle 
positions \cite{reichhardt2015softening}. However, the effect of friction on 
the elastic softening \cite{jia2011elastic,wildenberg2013} is absent in these 
frictionless systems. 

Here we address this issue by simulating the effects of resonance frequency 
softening in confined 3D frictional bead packings under an applied dynamic (AC) 
drive.  We drive the confined granular packing at larger and larger amplitudes 
to explore the nonlinear response of the material that provide clues to the 
transition from a solid state to a fluid state where the mobility of particles 
becomes important, nevertheless having a mean-square displacement smaller than 
the particle size.  Note that the granular packing does not flow in our 
simulations due to the absence of macroscopic shear, but merely approaches the 
fluid state with significant elastic softening and particle position 
rearrangement \cite{dAnna2001jamming,jia2011elastic}. In the next section, we 
present the numerical model and the protocol of tests and then we show the 
simulation results. All these findings will be analysed using the mean-field 
approach.


\begin{figure}
\centering
\includegraphics[width=1.0\columnwidth,clip]{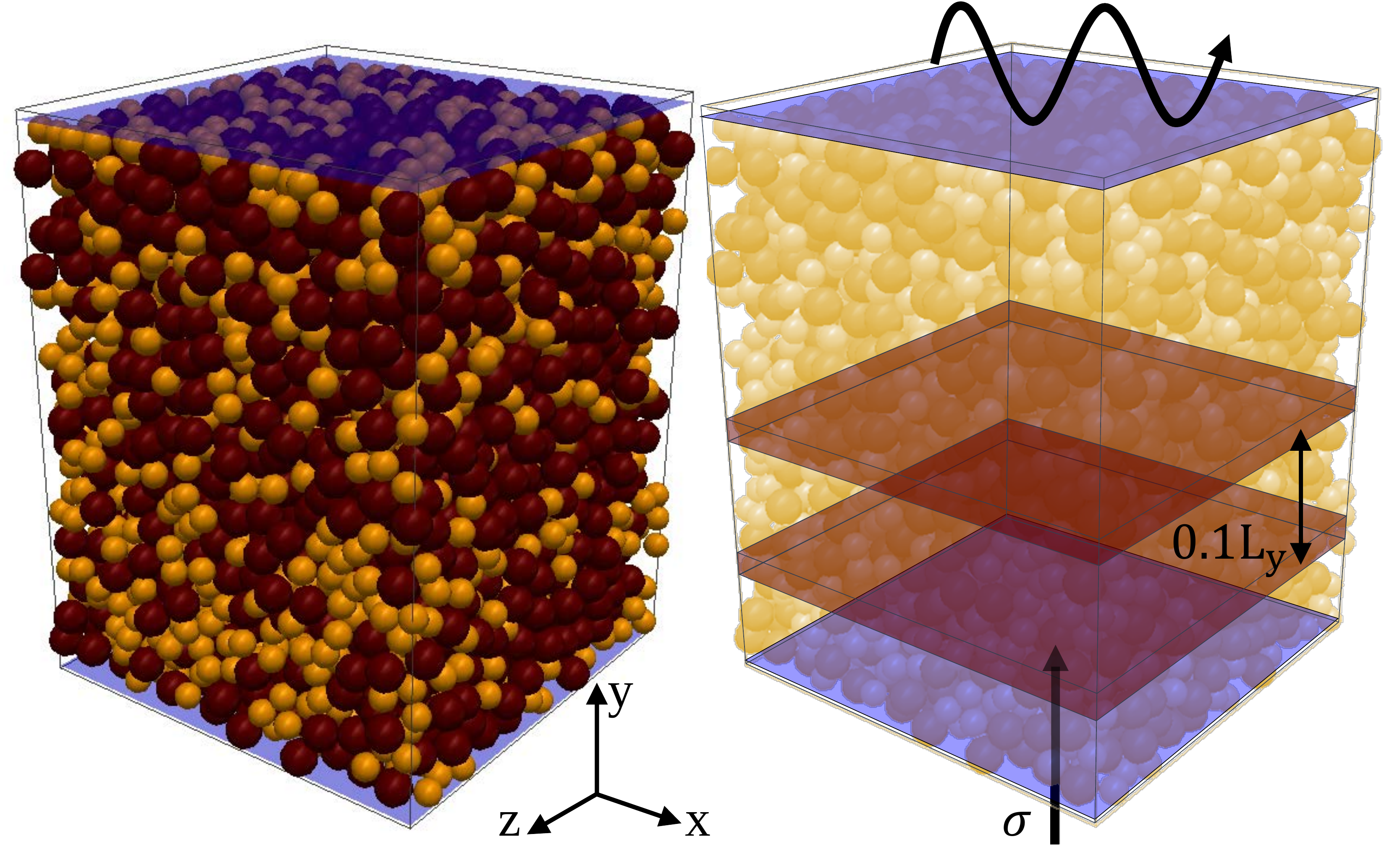}
\caption{(left) Representation of a sample, with large particles in red. The 
packing is enclosed between two walls along the y-direction. (right) The two 
red sections enclose a local region of the sample, where wall effects are 
minimal. The perturbation are movements of the top wall while the bottom wall 
adjusts to maintain constant loading stress.}
\label{figsetup} 
\end{figure} 

\section{\label{methodologysection} Numerical model and methodology}

We apply the Discrete Element Method \cite{cundall1979discrete} (LIGGGHTS 3.4 
\cite{kloss2011liggghts}) to simulate the mechanical behavior of 3D glass bead 
packings under static and oscillatory loading stress. The Hertz contact law is 
used to reproduce normal elastic interactions between elastic spheres and an 
elastofrictional model is implemented to describe the tangential interaction. 
The collision process is also included for the viscoelastic damping. 

\subsection{Intergrain forces}

The total force $\bm{F}_i$ acting on a particle $i$ is the sum of the contact 
forces with interacting particles $j\in J$ and viscosity damping. Specifically, 
it is the sum of a normal force $\bm{F}_n$ based on the Hertz theory, and a 
tangential force $\bm{F}_t$ based on an approximation of the Mindlin model 
\cite{johnson1985normal}: 

\begin{equation}%
\begin{aligned}%
\bm{F}_i &= \sum_{j\in J}\left(k_n \delta n_{ij}^{3/2} \bm{\hat n}_{ij} - 
\gamma_n  \bm{v}_{ij}\cdot \bm{\hat n}_{ij}\right) \\ &+\sum_{j\in J}\left(k_t 
\delta n_{ij} \delta t_{ij} \bm{\hat t}_{ij} - \gamma_t  \bm{v}_{ij}\cdot 
\bm{\hat t}_{ij}\right) - \gamma_{a} \bm{v}_i
\end{aligned}%
\label{allforces}\end{equation}%
where $\delta n_{ij}$ and $\delta t_{ij}$ are the normal overlap and relative 
tangential displacement between particles $i$ and $j$.  $\bm{\delta t}_{ij}$ is 
truncated to fulfil $\|\bm{f}_{t}\| < \mu_s \|\bm{f}_{n}\|$ where $\mu_s$ is 
the coefficient of the Coulomb friction, $\bm{f}_{n}$ is the normal force at 
one contact and $\bm{f}_{t}$ is the tangential force. Particles over threshold 
are modelled as slipping against each other. $\bm{\hat n}_{ij}$ and $\bm{\hat 
t}_{ij}$ are the normal and tangential unit vectors of each contact $ij$. 
Finally, $\bm{v}_i$ is the velocity of particle $i$  while $\bm{v}_{ij} = 
\bm{v}_j - \bm{v}_i$. $\gamma_a$ is the viscous damping coefficient for 
numerical stabilisation. 

Provided with the Young's modulus $Y_g$, shear modulus $G_g$ and loss 
coefficient $\beta_g$ of the grains, the tangential and normal elastic 
constants are: 

\begin{align}%
k_n &= \frac{4}{3} Y_g \sqrt{R^*} \\
k_t &= 8 G_g \sqrt{R^*}  %
\end{align}%
and the viscoelastic parameters related to collision losses: 
\begin{align}%
\gamma_n &= 2 \beta_g (R^* \delta n_{ij})^{1/4} \sqrt{\frac{5}{3}m^* Y_g}   \\
\gamma_t &= 4 \beta_g (R^* \delta n_{ij})^{1/4} \sqrt{\frac{10}{6}m^* G_g}%
\end{align}%
The effective radius $R^*$ and mass $m^*$ depend on the respective radii $R$ 
and masses $m$ of the two particles: $1/R^* = 1/R_i + 1/R_j$ and $1/m^* = 1/m_i 
+ 1/m_j$.

The equation governing rotational velocity is:
\begin{equation}%
I_i\dfrac{d\bm{\Omega}_i}{dt} = \bm{r}_{i} \times \bm{F}_{t,i} + \bm{T}_{i,j}
\end{equation}%
with $I_i$ the moment of inertia, $\bm{r}_{i}$ the vector from the center of 
the particle $i$ to the contact point, $\bm{F}_{t,i}$ the tangential component 
of the force  exerted on the particle $i$ and a torque at each contact: 
\begin{equation}%
\bm{T}_{i,j} = \mu_r k_n \delta n_{ij} R_i 
\dfrac{\bm{\Delta\Omega}_{ij}}{\|\bm{\Delta\omega}_{ij}\|}
\end{equation}%
with $\mu_r$ the rolling resistance and  $\bm{\Delta\Omega}_{ij} = 
(R_i\bm{\Omega}_i + R_j \bm{\Omega}_j)/(R_i+R_j)$ is the relative angular 
velocity \cite{goniva2012influence}.

Energy is basically dissipated via frictional damping in sliding and, in 
collisonal interactions, via the coefficient $\beta = 
ln(e)/\sqrt{ln^2(e)+\pi^2}$, with $e$ the restitution coefficient.

\begin{figure}
	\centering
	\includegraphics[width=1.0\columnwidth,clip]{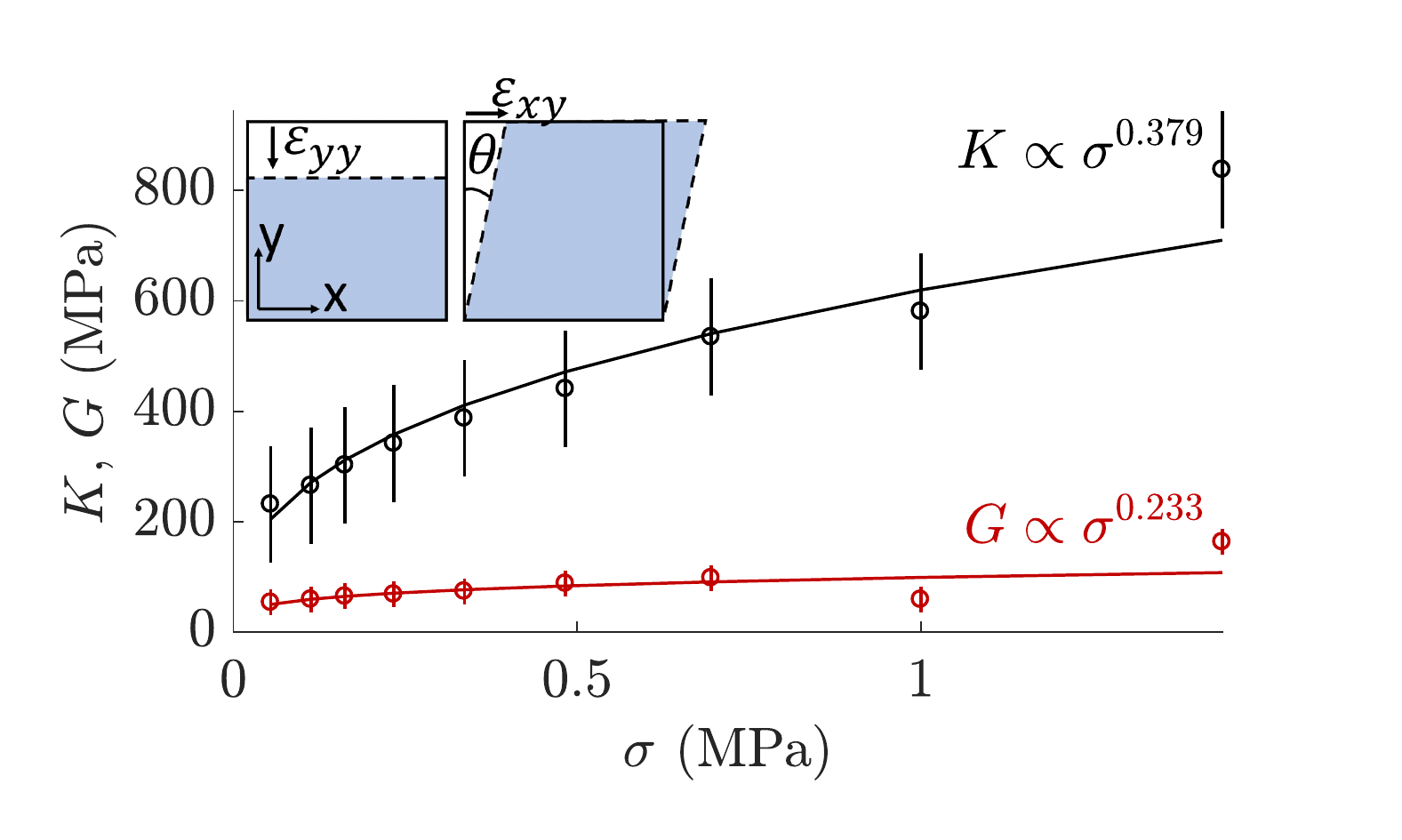}
	\caption{Elastic and shear moduli $K$ and $G$ of the granular samples as a 
	function of loading stress $\sigma$. 
		The inset shows the compression and shear protocols. }
	\label{figmoduli} 
\end{figure} 

\subsection{Granular sample preparation}

3D samples are created by randomly filling particles of radius $r_1$ or $r_2$ 
into a box with dimensions $(l_x,l^0_y,l_z)= 
(10^{-2},2.5\cdot10^{-2},10^{-2})\text{ m}$ (See Fig. \ref{figsetup}). 
Particles have a density $\rho_0 = 2500\text{kg/m}^3$. The $y$-axis is vertical 
(with $y=0$ at the bottom) and periodic boundary conditions are applied in the 
x- and z-directions. Two walls enclose the system in y-direction. These walls 
can be moved via a fixed function or servo-controlled, in which case an 
algorithm will move the wall with a maximal velocity $v_{max}$ to maintain a 
constant loading stress. 

A variable velocity from $v_{max} = 6\text{ m/s}$ to $0$ is applied to the 
bottom of the box to compact the particles while the top wall moves to reach a 
static loading stress $\sigma$. After a period $\Delta t_1 = 0.2\text{ s}$, the 
bottom wall moves with a sinusoidal vertical movement $A \sin(\omega t)$ with 
$\omega=1.26\cdot 10^5 \text{ rad/s}$ for a period $\Delta t_2 = 2\text{ s}$ to 
perturb the system and bring it closer to equilibrium. The amplitude of 
perturbation $A = 5\cdot10^{-5}\cdot l_y$ with $l_y$ the sample height when 
perturbation starts. Preliminary simulations showed that this value drives 
rearrangements and compaction in the samples. Finally the bottom wall is 
stopped and the packing is allowed to evolve and relax during period $\Delta 
t_3 = 7.8\text{ s}$, to reach a sample height of $l_y \approx 
1.2\cdot10^{-2}\text{ m}$ with about $4100$ particles. Granular samples are 
confined by the loading stress $\sigma$, set at $10$ logarithmically spaced 
values between $\sigma= 10\text{ kPa}$ and $\sigma= 1438\text{ kPa}$. The 
resulting static strain $\varepsilon_0$ is estimated as $\sim 10^{-4}-10^{-3}$ 
with a limited precision at low values depending on the test protocol.  

All simulations are run with $\gamma_a = 10^{-7}\text{ kg/s}$ (air effect at 
room temperature) and timestep $dt = 8\cdot10^{-8}\text{ s}$. Particles have 
contact force parameters of  $Y_g = 65\text{ GPa}$, $\nu_g = 0.25$ ($K_g = 
43.33 \text{ GPa}$), $\mu_s = 0.22$, $\beta_g = 0.0163$ and $\mu_r = 0.01$, and 
their radii are $r_1 = 3\cdot10^{-4}\text{ m}$ and $r_2 = 4\cdot10^{-4}\text{ 
m}$. $60\%$ of the sample mass is made of particles with radius $r_1$ and 
$40\%$ with radius $r_2$.

\begin{figure*}
	\centering
	\includegraphics[width=1.0\textwidth,trim={1.6cm 0 3cm 
	0},clip]{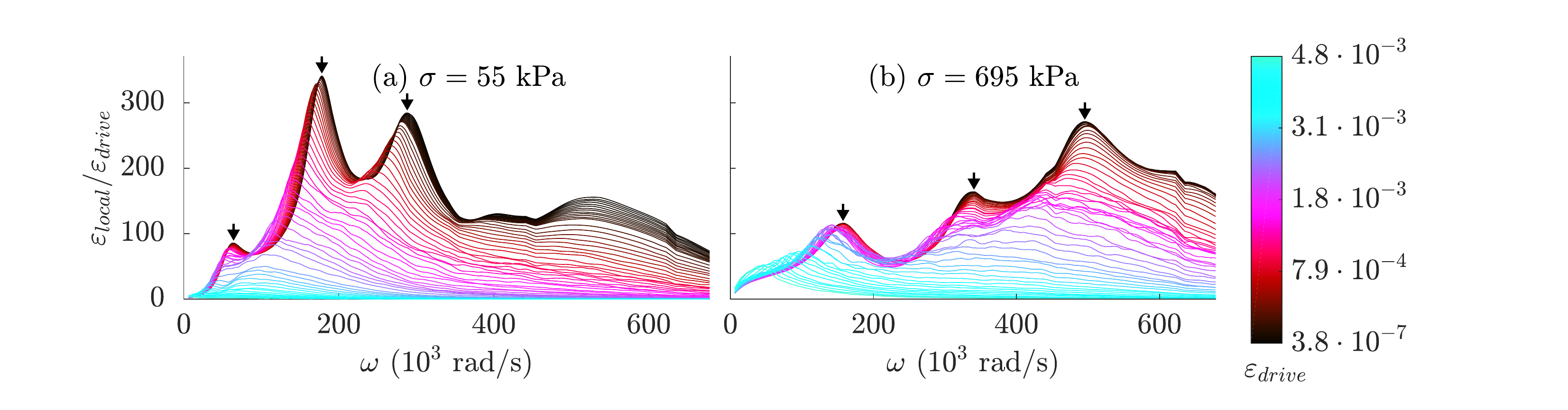}
	\caption{Two series of frequencies sweeps obtained for $\sigma = 55\text{ 
	kPa}$ (a) and $\sigma = 695\text{ kPa}$. Color represent the drive strain 
	of each sweep. Harmonics are visible. Resonant frequencies decrease and 
	peak broaden as drive strain increase. As shown in Fig. \ref{figvertical}, 
	each figure correspond to a different boundary condition. }
	\label{figsweeps} 
\end{figure*}

\subsection{Tests under quasistatic compression and shear}

The granular samples are compressed by uniaxial load or sheared at constant 
volume to determine their bulk and shear moduli, respectively (see figure 
\ref{figmoduli}, inset). The driving incremental strains (cycle) are $\Delta 
\varepsilon \sim \cdot10^{-6}$.

During a compression test, the bottom wall remains fixed while the top wall 
moves downward at constant velocity $v_{lid}$ to make a small cycle around the 
confining stress $\sigma$. The top wall first moves down at velocity $v_{lid}$ 
over a period $\Delta t_{compr}$, then up at $v_{lid}$ for $2\cdot \Delta 
t_{compr}$, then back down at $v_{lid}$ for $\Delta t_{compr}$. The compressive 
strain $\varepsilon_{yy} = (l_y-l_y^0)/l_y^0$ (the deviation of sample height 
$l_y$ from its initial value $l_y^0$) and compressive stress $\sigma_{yy}$ on 
the top wall are recorded. The stress for an isotropic packing is given as: 
\begin{equation} \sigma_{yy} = \dfrac{3 Y (1-\nu)}{(1+\nu)} \varepsilon_{yy} 
\label{myeq1} \end{equation} 
with Young's Modulus $Y$ and Poisson ratio $\nu$ of the granular medium. 

A shear test is performed by imposing the wall at fixed positions along the 
y-axis, while the top wall is sheared along the x-axis. The wall moves at 
velocity $v_{lid}$ in the x-direction for a time $\Delta t_{compr}$, then moves 
back at $v_{lid}$ for $2\cdot \Delta t_{compr}$, and finally at $v_{lid}$ for 
$\Delta t_{compr}$. The shear stress $\sigma_{yx}$ and the strain 
$\varepsilon_{yx} = (x_{lid}-x_{lid}^0)/l_y^0$ are recorded during the 
simulation. Assuming isotropic linear elastic behavior, the stress-strain 
relation is given as:
\begin{equation} \sigma_{yx} = 2G \varepsilon_{yx} \label{myeq2}
\end{equation} 
Equations \eqref{myeq1} and \eqref{myeq2} constitute a system of two equations 
with two unknowns: the Young's Modulus $Y$ and Poisson ratio $\nu$ of the 
granular material. The bulk and shear moduli are given by $K=Y/3(1-2\nu)$ and 
$G=Y/2(1+\nu)$. The moduli were determined using the full cycle, with 
negligible hysteresis observed during the loading-unloading.

Compression and shear tests were both performed at a velocity $v_{lid} = 
10^{-7}\text{ m/s}$ for $\Delta t_{compr} = 0.2 \text{ s}$. The highest 
velocity ensuring quasi-static measurements is computed for $I = 
2\dot{\gamma}r_{min}/\sqrt{\sigma_{min}/\rho} < 10^{-3}$, with $r_{min} = 
3\cdot10^{-4}\text{ m}$, $\sigma_{min} = 10\text{ kPa}$, $\rho \approx 
1400\text{ kg/m$^3$}$. The strain rate $\dot{\gamma} = v_{lid}/l_y$ implies 
$v_{lid} < 5.79\cdot10^{-2}\text{ m/s}$ for our samples.

\subsection{Tests under compressional vibration}

A probing layer inside the sample with y-coordinate comprised between $0.2\cdot 
l_y$ and $0.4\cdot l_y$
has been selected (to avoid the wall effect) to study the dynamic strain of the 
granular packing $\varepsilon_{local}$ to an applied vibration. The system is 
driven for $N(=100)$ periods at each drive frequency so that it reaches 
quasi-steady state and data are averaged over the last $N' = 60$ periods and 
recorded. Measurements are made $50$ times per period.

Frequency sweeps are performed in the granular packings under different 
confining stress $\sigma$, maintained by the bottom wall. The applied vibration 
is introduced by vertically moving the top wall with $A_{drive}\sin(\omega t)$ 
for $N$ periods. The frequency during sweeps is increased from $\omega= 
6.2\cdot10^3 \text{ rad/s}$ to $\omega= 6.2\cdot 10^{5} \text{ rad/s}$ by steps 
of $1.9\cdot10^{2} \text{ rad/s}$. The drive amplitude $A$ ranges from 
$5\cdot10^{-8}$ to $6\cdot10^{-5}\text{ m}$. The dynamic strain is computed as 
the amplitude $A_{drive}$ divided by the average height of the system $\langle 
l_y\rangle$ : $\varepsilon_{drive} = A_{drive}/\langle l_y\rangle$,varying 
accordingly from $4.13\cdot10^{-6}$ to $4.95\cdot10^{-3})$ 

For measuring the resonance response of the sample, we investigate the dynamic 
strain of a local layer away from the nodes of the standing waves. Here, the 
local strain $\varepsilon_{local}$ is computed as follows. First the 
instantaneous strain is determined as the relative difference between the local 
layer height $l(t)$ and its average $l_{\omega_i}$ over the last $N'=60$ 
periods of the system driven at $\omega_i$: 
\begin{equation} \varepsilon_{_{\omega_i}}(t) = \dfrac{l(t) 
-l_{\omega_i}}{l_{\omega_i}} \end{equation}
The strain $\varepsilon_{_{\omega_i}}(t)$ is composed of a slowly-evolving 
component due to the relaxation of the specimen and a component oscillating at 
drive frequency. We remove the slowly-evolving component by subtracting a 
moving-average window with a length equivalent to one period of the system. 
Finally we fit the amplitude and phase of the sinus $\varepsilon_{local} sin 
(\omega t + \phi)$ on the remaining strain oscillation.

\subsection{Slipping contact ratio and coordination number}

In addition to the dynamic strain, we also studies the Slipping Contact Ratio 
$SCR(t)$ of the probe layer which is the ratio between contacts 'slipping' in 
regard to each other (algorithmically when the Coulomb threshold is applied to 
the tangential force) to all contacts in the probing layer. The average 
$\langle SCR \rangle$ corresponding to a frequency $\omega_i$ is obtained over 
the last $N'$ periods of $SCR(t)$ driven at $\omega_i$. Similarly, the 
coordination number $Z_c(t)$ is the average number of contacts of particles in 
the probing layer. $\langle Z_c \rangle$ is the average of $Z_c(t)$ over the 
last $N'$ periods. In this work, we investigate the values of 
$\varepsilon_{local}$, $\langle Z_c \rangle$ and $\langle SCR \rangle$ in the 
same probing layer as a function of the driving frequency $\omega$. 


\begin{figure}
\centering
\includegraphics[width=1.0\columnwidth,clip]{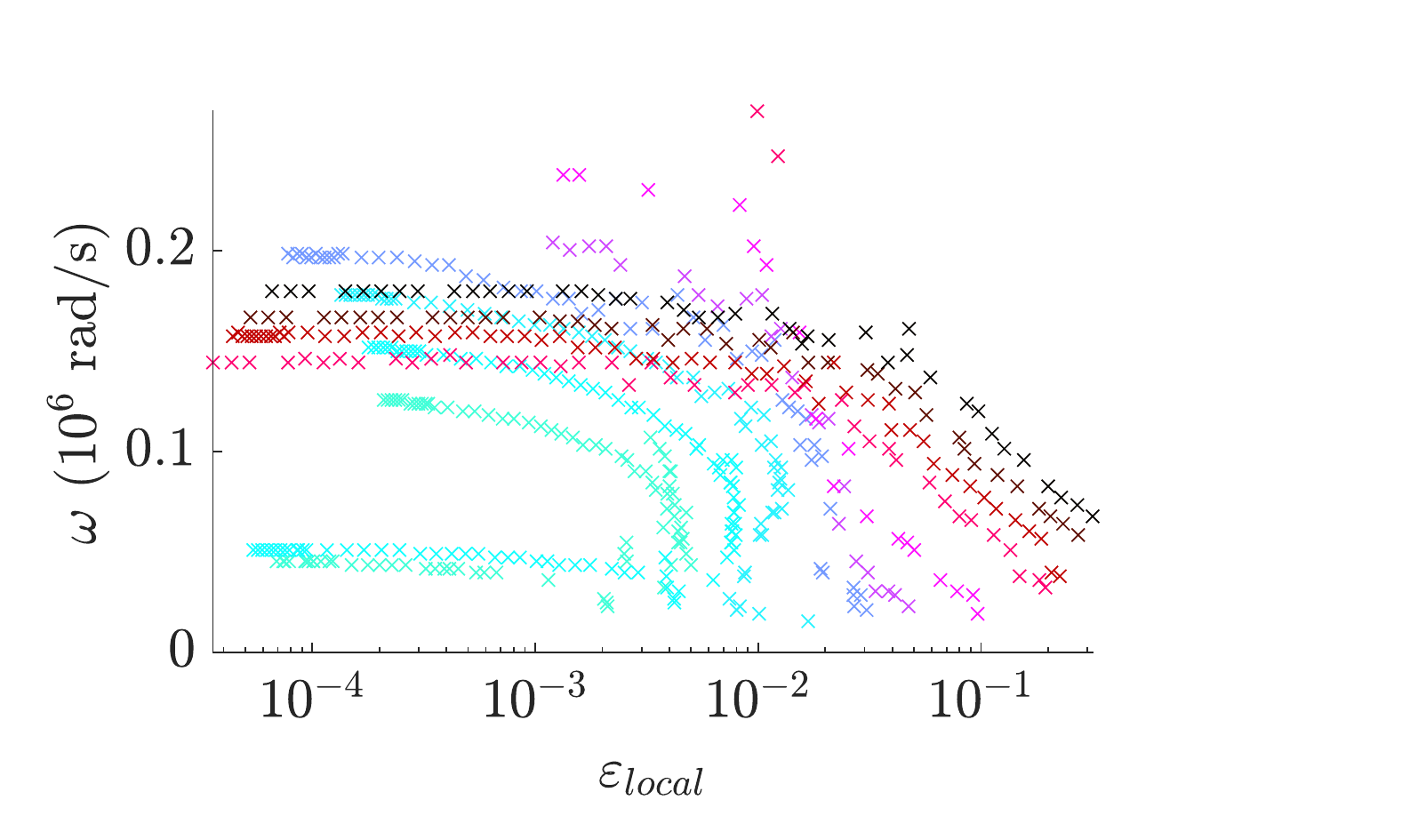}

\includegraphics[width=1.0\columnwidth,clip]{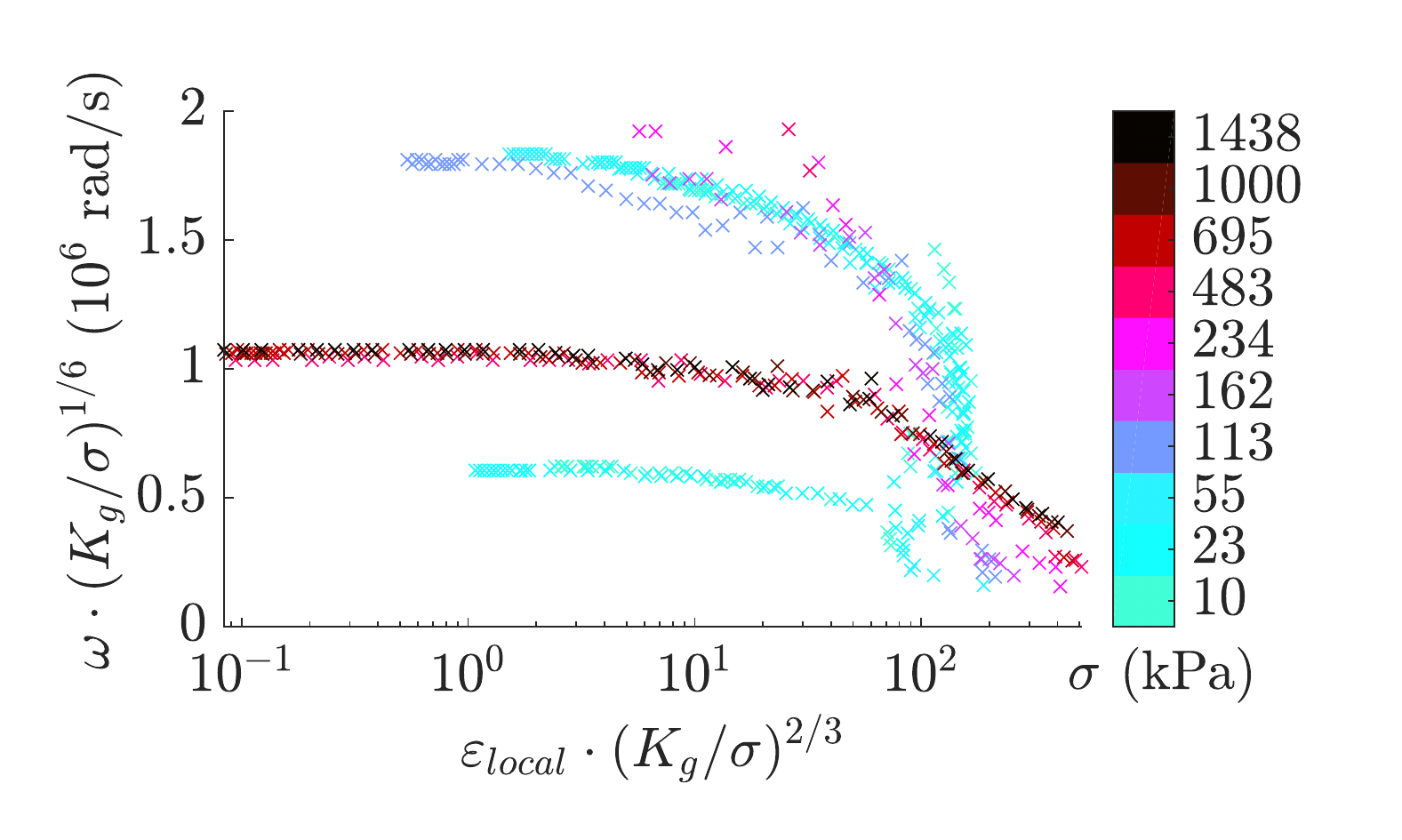}
\caption{The first modes are extracted from the frequency sweeps (Fig. 
\ref{figsweeps} and reported in (a) as a function of local strain. When scaled 
according to predictions of Hertzian theory, resonances collapse onto primary 
curves (b). The two resonances correspond to different modes (Fig. 
\ref{figvertical}). Resonant frequency remains constant for low drive strain, 
then decreases above a threshold value. }
\label{figpeaks} 
\end{figure}

\section{\label{resultsection} Results and discussion}

\subsection{Bulk moduli}

Fig. \ref{figmoduli}) shows the bulk $K$ and shear $G$ moduli obtained by 
compression and shear tests in granular samples as a function of the confining 
stress $\sigma$. $K$ scales as $\sigma^{0.379}$ being close to the Hertz theory 
prediction with a dependency of $\sigma^{1/3}$ \cite{walton1987effective} 
whereas $G$ scales as $\sigma^{0.233}$ with an exponent lower than those found 
in other simulations with imposed confining pressure \cite{makse2004granular, 
agnolin2007III, wyart2005effects}. As shown below, the value of bulk modulus 
found here are in good agreement with the compressional sound velocity deduced 
from our main resonance simulation. We also observe a strong dependence of the 
elastic on the coordination number and packing fraction of the samples 
\cite{agnolin2007III}. In the present samples, the coordination number ranges 
from $4.38$ to $4.95$, and the packing fraction from $0.60$ to $0.62$, in the 
range expected for an isotropic packing of frictional particles 
\cite{agnolin2007I}.

\begin{figure}
\centering
\includegraphics[width=1.0\columnwidth,clip]{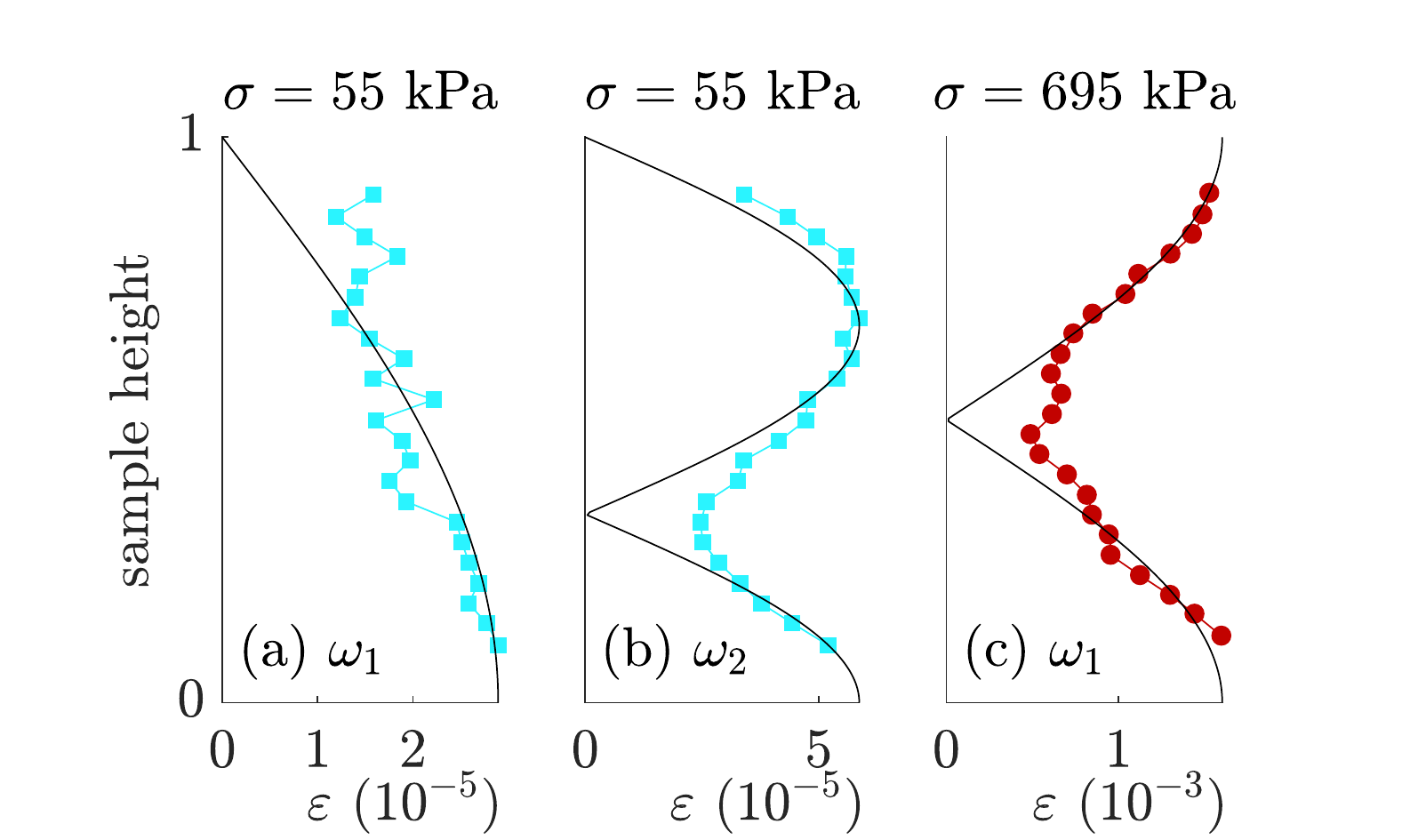}
\caption{Each figure shows the strain profile of the sample vibrated at 
resonance. The first and second modes (a) and (b) of a low stress sample show a 
closed boundary at the top and open at the bottom. In the high stress sample 
(c), both boundaries are open.   }
\label{figvertical} 
\end{figure} 

\subsection{Resonance, Hertzian scaling and softening}

Figure \ref{figsweeps} shows two series of frequency sweeps performed for low 
$\sigma = 10\text{ kPa}$ and high $\sigma = 695\text{ kPa}$, respectively. Both 
figures show the results for a series of drive strains from 
$\varepsilon_{drive} = 3.8\cdot10^{-7}$ to $4.8\cdot10^{-3}$ (the highest value 
for which a peak could be distinguished for all samples). At low drive strain, 
the resonance frequency does not decrease, but the system exhibits harmonics 
(shown with arrows on Fig. \ref{figsweeps}) as found in the experiments 
\cite{brunet2008}.	
As the drive strain increases, resonance frequencies decrease and resonant 
peaks broaden and start to merge with each other. 

We plot these frequency shifts in Fig. \ref{figpeaks}. The resonances $\omega$ 
satisfying $\omega < 2.5\cdot10^{5}\text{ rad/s}$ are extracted detecting and 
fitting each peak with a third-order polynomial, and reported in Fig. 
\ref{figpeaks}(a). Samples at high loading stress, above $480\text{ kPa}$ show 
a clear linear regime at low local strain $\varepsilon_{local}$ followed by a 
softening. This thresholding was selected because the two first modes at low 
loading stress merge as drive strain increases and, at high loading stress and 
high drive strain, only the first mode remains. The Hertz theory, as 
implemented in equation \eqref{allforces}, predicts a contact force 
proportional to the overlap between two particles $f_n \propto \delta^{3/2}$. 
If our samples were to follow the mean-field approach based on the Hertzian 
contact, the resonance $\omega$ would scale as $\sigma^{1/6}$  
\cite{goddard1990nonlinear} and the strain $\varepsilon$ as $\sigma^{2/3}$  
\cite{walton1987effective}. We apply this scaling and report the data on Fig. 
\ref{figpeaks}(b), where we observe that resonances collapse onto primary 
curves: two for low stresses and one for high stresses.

We note that resonant modes at $\sigma = 55\text{ kPa}$ and $\sigma = 695\text{ 
kPa}$ represent different boundary conditions. Fig. \ref{figvertical}(a,b) 
shows the local strain $\varepsilon$ vertical profile in a sample at $\sigma = 
55\text{ kPa}$ vibrated at its first and second modes $\omega_1$ and 
$\omega_2$. The data shows an approximately free boundary condition at the top 
wall with very small strain or stress, and a clamped-like boundary at the 
bottom wall with finite strain or stress. The first modes of such a system are 
$\lambda/4$ and $3\lambda/4$ (black line on (a) and (b)).  In contrast, the 
high loading stress samples $\sigma = 695\text{ kPa}$ shows clamped-like 
boundary conditions at both ends where the dynamic strain or stress is very 
small compared to the static strain or confining stress. The first mode, 
$\lambda/2$, and data are reported on Fig. \ref{figvertical}(c). Data in all 
three cases follow the predicted modes. The deviation on the boundary and at 
the antinode are presumably due to the inherent fluctuation in the granular 
material. Fig. \ref{figpeaks}(b) thus shows a Hertzian scaling of the resonance 
for two different boundary conditions. At low loading stress, two modes are 
observed which soften at higher local strain, and merge together. Data for high 
loading stress supports higher local strains and shows a clear linear regime at 
low local strain. Moreover, from Figs. 4a and 5c, we measure a resonant 
frequency $\omega \sim 1.5\cdot10^5 \text{rad/s}$ with a wavelength $\lambda = 
2 l_y$ which gives a longitudinal sound speed $c = \lambda\omega/2\pi = 500 
\text{ m/s}$. The bulk modulus can then be deduced by $K = \rho c^2 \sim 625 
\text{ MPa}$, which is consistent with the value derived from the static 
compressional test (Fig. \ref{figmoduli}).

\begin{figure}
\centering
\includegraphics[width=1.0\columnwidth,clip]{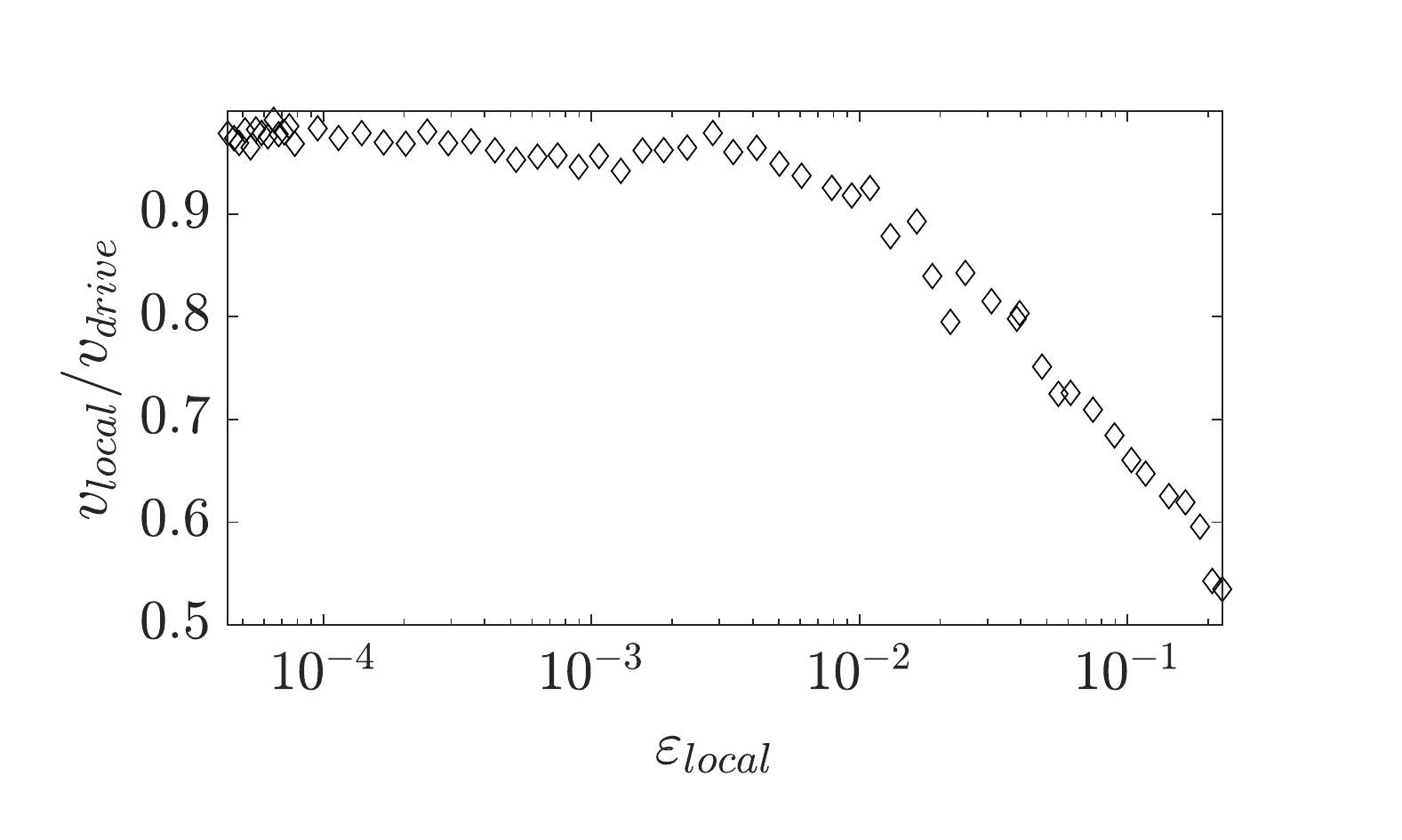} 
\caption{Ratio of the average particle velocity $v_{local}$ in the local layer 
to the maximal velocity of the wall $v_{drive}$ under $\sigma = 695\text{ 
kPa}$. At low strain, both velocities are roughly equal. At high strain, the 
local particle velocity decreases compared to drive velocity. }
\label{figvpvlid} 
\end{figure} 

We also investigate the transmission of the kinetic energy into the granular 
system from the driving force, by comparing the average particle velocity 
$v_{local}$ in the probing layer and the driving wall velocity $v_{drive}$. The 
particle velocity is obtained from the relation $E_{kin} = (1/2) 
M_{g}v_{local}^2$, with $E_{kin}$ returned by simulations and the packing mass 
$M_{g} = 1.15\cdot10^{-3}\text{ kg}$ for $\sigma = 695\text{ kPa}$. The maximal 
driving velocity of the wall is the product of the resonant frequency and the 
displacement amplitude: $v_{drive} = \omega\cdot A$. Fig. \ref{figvpvlid} shows 
that at low strain (linear regime) the ratio of the particle velocity to the 
driving is almost constant and equal to $1$, but at higher strain it decreases 
to about $0.5$. Furthermore we observe a striking similarity between this ratio 
and the resonance frequency decrease with increasing dynamic strain (Fig. 
\ref{figpeaks}(b) and Fig. \ref{figfriction}(a)). This behaviour can be 
qualitatively understood in terms of the coefficient of amplitude transmission 
which depends on the acoustic impedance and accordingly the sound velocity of 
the granular sample \cite{wildenberg2013}: the lower the sound velocity (due to 
elastic softening), the lower the transmission coefficient.


\begin{figure}
\centering
\includegraphics[width=1.0\columnwidth,clip]{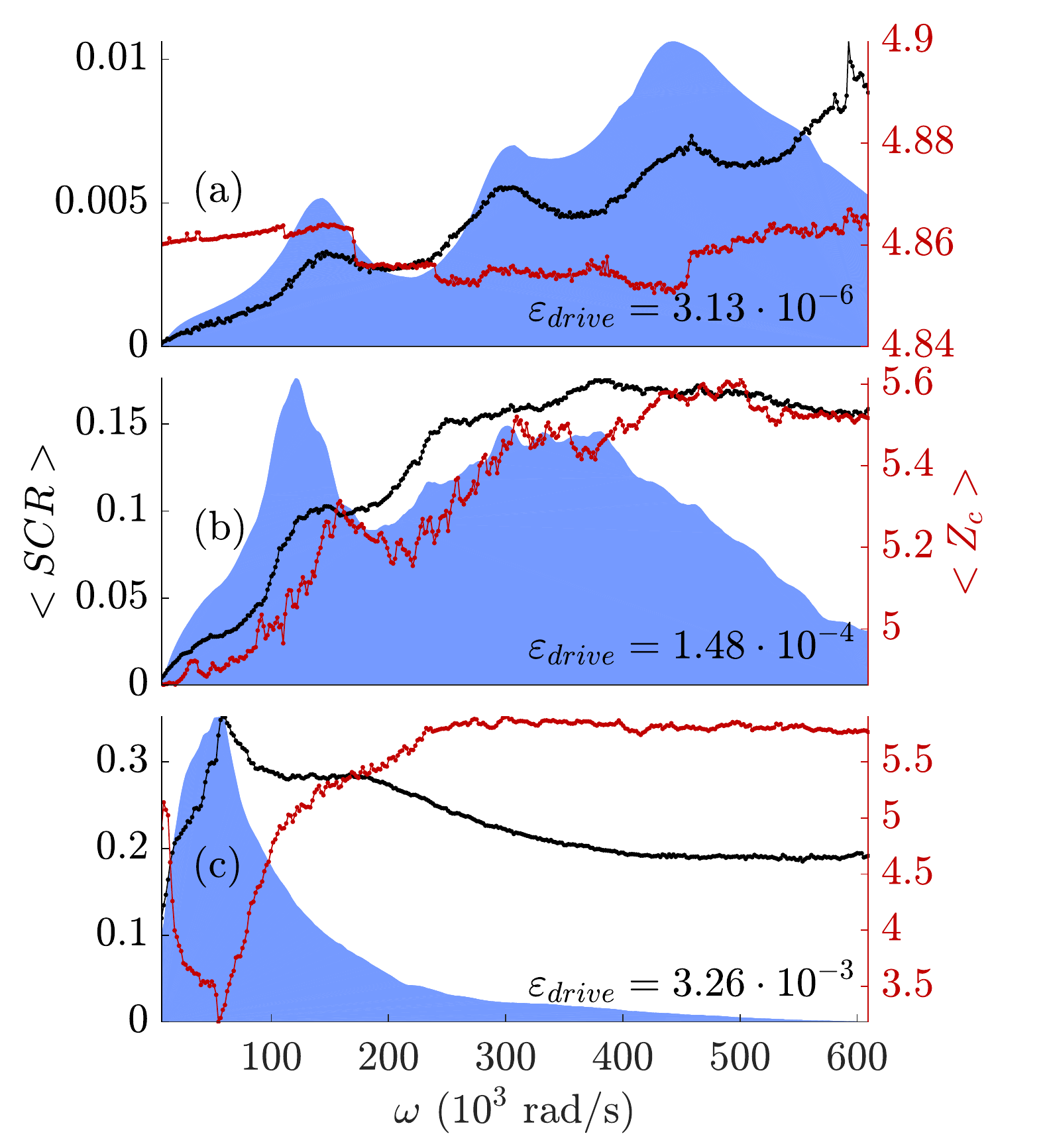} %
\caption{Coordination number $\langle Z_c \rangle$ (right-hand vertical axis), 
Slipping Contact Ratio $\langle SCR \rangle$ (left-hand vertical axis) and 
normalized frequency spectrum (blue) in the linear (a), transition (b) and 
nonlinear (c) regions of the drive strain at loading stress $\sigma= 695\text{ 
kPa}$. Shifts in $\langle Z_c \rangle$ are below $0.01$ for low drive strain. 
In the transition regime, Its average value increases as a result of 
perturbation up to a saturation value of $5.71$. $\langle SCR \rangle$ 
increases peaks at resonant frequencies. In the nonlinear regime the system 
softens, and the coordination number decreases at resonance. Left axis shows 
the variation range for each case.  }
\label{figZc} 
\end{figure}


\subsection{Changes of coordination number and slipping contact ratio upon 
applied vibration}

To understand the responses of granular samples to the applied vibration, we 
investigate simultaneously the structural changes of the contact networks. More 
specifically, we study the coordination number and slipping contact ratio on 
the grain scale in our samples. Fig. \ref{figZc} shows three frequency sweeps 
performed on a sample at $\sigma = 695\text{ kPa}$. The blue outline represents 
a scaled profile of the frequency sweep. In black is reported the percentage of 
slipping contacts $\langle SCR\rangle$ (left axis) and in red the average 
coordination number $\langle Z_c \rangle$ (right axis). In the linear regime 
(a), the coordination number remains constant at $\sim 4.86$, with variations 
due to small movements in the sample. The slipping contact ratio peaks at 
resonance, and progressively increases from $0$ to $1\%$ over the frequency 
sweep, showing a progressive mobilisation of contacts. At higher drive strain 
(b), the coordination number peaks around the first resonant mode and spans the 
range $4.8$ to $5.5$ over the frequency sweep. The slipping contact ratio peaks 
at the first resonant mode.  At higher drive strain (c), the coordination 
number decreases at resonance frequency. The slipping contact ratio shows a 
moderate peak at resonance, than decreases after reaching $20\%$ of mobilised 
contacts. Results in (b) and (c) show lasting changes in the grain arrangement 
(coordination number), and dynamic changes in the sample (slipping contact 
ratio). 

\begin{figure}
\centering
\includegraphics[width=1.0\columnwidth,clip]{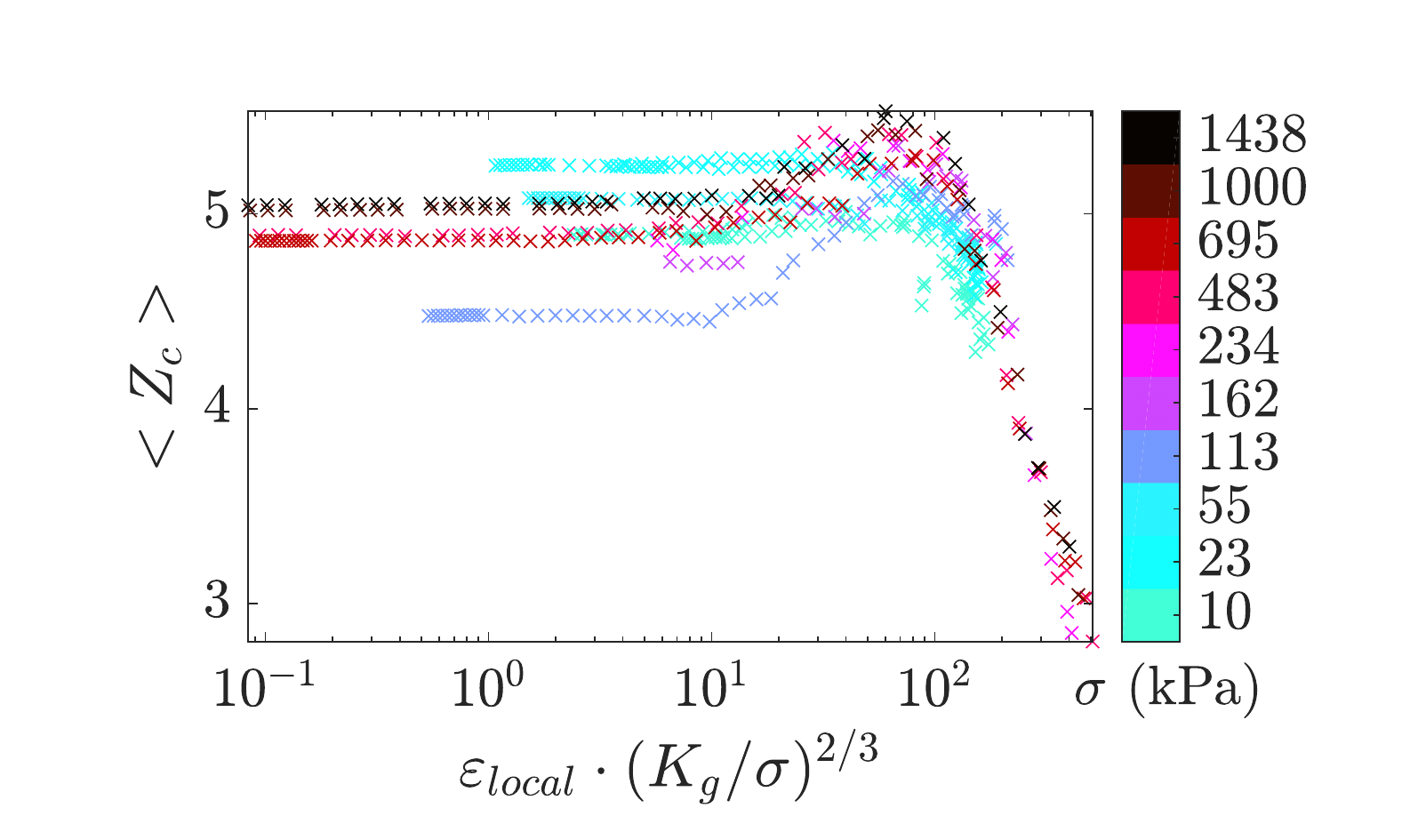} 
\caption{Coordination number at resonance for the first mode of the different 
packings. In all cases the coordination number is constant at low strain and 
decreases at high strains. }
\label{fig3} 
\end{figure}

Fig. \ref{fig3} reports the average coordination number at resonance for 
different loading stresses and local strains. In all cases, the average is 
constant at low strain and decreases at high strain. High loading stresses 
reach higher local strain values and we therefore use the sample at  $\sigma = 
695\text{ kPa}$ to study the behavior of the system. Variations in initial 
coordination number (value at low strain) are due to randomness in the packing 
generation. 

\begin{figure}
\centering
\includegraphics[width=1.0\columnwidth,trim={0 0.45cm 0 
0.45cm},clip]{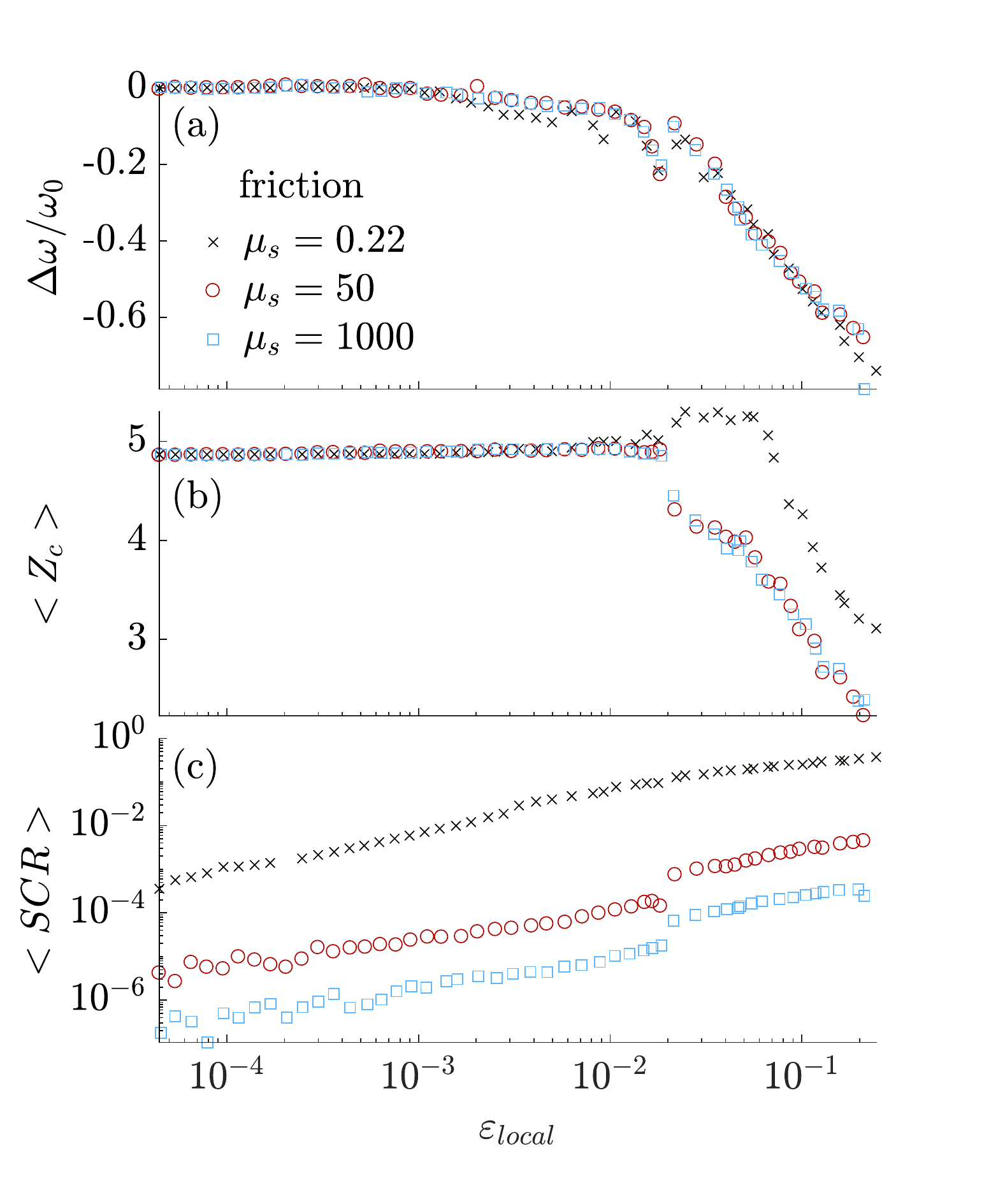} 
\caption{(a) resonant frequency of sample at $\sigma = 695\text{ kPa}$ for a 
range of local strains and three friction coefficients $\mu_s$. $\mu_s$ has no 
impact on the softening or the transition from linear to nonlinear. (b) 
Coordination number $\langle Z_c \rangle$ and (b) slipping contact ratio 
$\langle SCR \rangle$ at resonance for the friction coefficients. The behavior 
of $\langle Z_c \rangle$ is similar and, at high friction, closely follows the 
shift in resonant frequency of (a). $\langle SCR \rangle$ is reduced by $10^3$ 
orders of magnitude and thus does not contribute to the softening observed in 
(a). }
\label{figfriction} 
\end{figure}

In order to test the effect of slipping contacts on the softening, the resonant 
frequencies of the sample at $\sigma = 695\text{ kPa}$ are reported for 
different values of the friction $\mu_s$ in Fig. \ref{figfriction}(a). The 
linear and nonlinear regime are identical whether contacts can slip 
($\mu_s=0.22$) and almost not ($\mu_s = 50$ and $\mu_s = 1000$). Fig. 
\ref{figfriction}(b) shows the coordination number for the three friction 
coefficients. The linear regime spans the same range of strain, and all three 
systems show a decrease of coordination number in the nonlinear regime. In the 
case of high friction, curves in Fig. \ref{figfriction} (a) and (b) have a 
similar shape. Fig. \ref{figfriction} (c) reports the slipping contact ratio 
$\langle SCR \rangle$ as a function of local strain $\varepsilon_{local}$. For 
a specific friction coefficient $x_\mu$, the slipping contact ratio increases 
with local strain. It is scaled down by three orders of magnitude between 
$\mu_s = 0.22$ and $\mu_s = 1000$, but without any clear correlation with the 
frequency softening. The tangential component of the interaction force has no 
or limited impact on the material softening, in contrast with the previous 
experimental observations where the microslip between the solid beads leads to 
the softening of contact stiffness and consequently the effective material 
softening (\cite{jia2011elastic}).


\section{Concluding remarks}

We form granular systems, subject to a normal Hertzian force and a tangential 
elasto-frictional force, in mechanical equilibrium at a sequence of confining 
stresses $\sigma = K \varepsilon_0$, where $K$ is the effective bulk modulus of 
granular samples (Fig. \ref{figmoduli}) and $10^{-5}\leq \varepsilon_0 \leq 
2\cdot10^{-3}$ . We subject these systems to dynamic testing using AC drive 
with amplitude $A$ at frequency $\omega$. The output of this dynamic testing at 
$10^{-7}\leq \varepsilon_{drive} \leq 2\cdot10^{-3}$ is the resonance frequency 
$\omega$ and the local AC strain $\varepsilon_{local}$ as a function of $A$ or 
$\varepsilon_{drive}$. We find to good approximation that the scaled resonance 
frequency is a universal function of the scaled strain, with the scaling 
appropriate to a Hertz contact system, $\omega \propto \sigma^{1/6}$, despite 
of important nonaffine motion of the particles with mean-square displacement 
being smaller than the particle size.

At low drive amplitudes, $\varepsilon_{drive} < 10^{-4}$, the dynamic response 
is linear and the effective elastic constant deduced from the resonance 
frequency is independent of $A$. At high drive strain $\varepsilon_{drive} \sim 
10^{-3}$ being close the static strain by the (high) confining stress, there 
are large departures from linearity. There the resonance frequency decreases as 
drive amplitude increases. These departures are taken to represent a reduction 
in dynamic modulus, termed 'softening'. The softening is a manifestation of the 
path toward unjamming transition; one infers that at slightly larger drive 
levels, the material will fluidize. The Hertz scaling extends into the AC 
amplitude domain in which we observe this softening. Throughout our exploration 
over ($S$,$A$)-space (to be detailed elsewhere), we monitor a number of 
quantities that could shed light on the microscopic properties of the system, 
e.g., the average coordination number $\langle Z_c \rangle$, the slipping 
contact ration $\langle SCR \rangle$ and the average particle velocity $v_p$. 

It is natural to ask if the slipping contact plays a role on the observed 
elastic softening. We repeated a number of calculations, primarily at $\sigma = 
695\text{ kPa}$, with the friction coefficient that controls 'slipping' set to 
values that include almost no slipping, and examined the slipping contact 
ratio. Even when the number of slipping contacts is reduced by several orders 
of magnitude we find no significant change in the nonlinear elastic response. 
The motion afforded by slipping contacts is not the responsible mechanism that 
contributes to the softening, as found in the experimental observations at 
strain amplitude of $10^{-6}$ due to the nonlinear tangential Mindlin contact 
\cite{johnson2005nonlinear, jia2011elastic}. Such contact softening mechanism 
is absent in the present numerical model, but our finding suggests another 
mechanism of softening likely related to the decrease of the coordination 
number due to the rearrangement of the particle position, driven by the applied 
AC amplitude in a manner similar to effective temperature characterized by 
non-affine motion of the particles \cite{dAnna2001jamming}.
The results are different from those found in 2D simulations where the particle 
positions are little changed under AC drive \cite{reichhardt2015softening}. The 
modulus reduction would be proportional to the amplitude of particle 
rearrangements which in turn is proportional to the amplitude of the AC drive. 
The modulus reduction is associated with increased fragility ultimately leading 
to fluidization, a phase transition not explored here. 

In summary we have applied DEM to study the behavior of a particle ensemble, 
driven progressively harder under resonance conditions.  We have done this at 
several applied loads.  We find a material softening manifested by the 
resonance frequency decrease with AC amplitude.  The coordination number is the 
most telling characteristic measured, showing the softening arises from the 
breaking and remake of contacts that are brought into existence and given 
mobility of the particles by the AC drive amplitude.  This approach could be 
valuably applied to the transition from the solid to the fluid states, which is 
approached in these simulations.  
\section*{Acknowledgements}%
Paul Johnson and Robert Guyer gratefully acknowledge the US Department of 
Energy Office of Basic Energy Science and support from EMPA. 
%
%
%


\end{document}